\documentstyle[preprint,aps,epsfig]{revtex}

\newcommand{\eq}{\begin{eqnarray}}
\newcommand{\en}{\end{eqnarray}}

\def \ie{{\it i.e.\,\,}}
\def \mate<#1|#2|#3>{\mbox{$\langle {#1}|\,{#2}\,|{#3}\rangle$}}

\begin{document}

\title{Ground-state baryon masses in\\ 
 the perturbative chiral quark model}
\author{T. \, Inoue,                                                   
V. \ E. \ Lyubovitskij,  
Th. \ Gutsche and 
Amand Faessler\vspace*{0.4\baselineskip}}
\address{
Institut f\"ur Theoretische Physik, Universit\"at T\"ubingen, \\
Auf der Morgenstelle 14,  D-72076 T\"ubingen, Germany 
\vspace*{0.3\baselineskip}}

\maketitle

\begin{abstract}
Mass differences of the flavor octet and decuplet 
ground-state baryons are studied in the perturbative chiral quark model. 
We present a way to understand the nontrivial spin- and flavor dependent
mass differences, where both pseudoscalar mesons and gluons play a
significant role.      
\end{abstract}

\vskip1cm


\vskip.5cm


\section{Introduction} 
In the na\"{\i}ve non-interacting quark model the masses of ground-state 
baryons are solely classified by their strangeness. Therefore, in this 
limit the masses of nucleon and $\Delta$, and also that of $\Lambda$ and 
$\Sigma$, are degenerate. Traditionally, this degeneracy is 
removed by introducing a hyperfine(spin-spin) 
interaction~\cite{Close:bt,Donoghue:dd} 
\eq
H_{\mbox{\footnotesize hyp}} 
= \frac12 \sum\limits_{i \, < \, j} {\cal H}_{ij} \, 
\vec s_i \cdot \vec s_j \, \delta^3(\vec r_i -\vec r_j)
\label{eqn:hyperint}
\en
where $\vec s_i$ is the spin operator acting on the $i$-th quark. 
Here, ${\cal H}_{ij}$ is the two-body quark coupling which includes 
a common color factor $-2/3$ and explicitly depends on the flavor of 
the constituent quarks through their masses $m_i$ and $m_j$: 
\eq
{\cal H}_{ij} \, \sim \, \frac{2}{3} \, \frac{1}{m_i \, m_j} \,. 
\en
The use of $SU(6)$ spin-flavor wave functions for the ground-state 
baryons $B$ leads to simple relations between the matrix elements 
$\mate<B|H_{\mbox{\footnotesize hyp}}|B>$,
which are the perturbative mass shifts due to the hyperfine interaction. 
Denoting the contribution from a non-strange quark pair as 
${\cal H}_{qq}$ (and similarly for strange quarks with $q$ replaced 
by $s$), in the isospin limit the masses of the ground-state baryons 
are composed as\cite{Donoghue:dd}
\begin{eqnarray}	
 \begin{array}{l}
  m_N         = 3 E_0            - \frac38 {\cal H}_{qq} \\
  m_{\Lambda} = 2 E_0 +   E_0^s  - \frac38 {\cal H}_{qq} \\
  m_{\Sigma}  = 2 E_0 +   E_0^s  + \frac18 {\cal H}_{qq} - \frac12 {\cal H}_{qs} \\
  m_{\Xi}     =   E_0 + 2 E_0^s  - \frac12 {\cal H}_{qs} + \frac18 {\cal H}_{ss}
 \end{array} ~~
 \begin{array}{l}
  m_{\Delta~}    = 3 E_0           + \frac38 {\cal H}_{qq} \\
  m_{\Sigma^\ast}= 2 E_0 +   E_0^s + \frac18 {\cal H}_{qq} + \frac14 {\cal H}_{qs} \\
  m_{\Xi^\ast}   =   E_0 + 2 E_0^s + \frac14 {\cal H}_{qs} + \frac18 {\cal H}_{ss} \\
  m_{\Omega~}    =         3 E_0^s + \frac38 {\cal H}_{ss} 
 \end{array}
\end{eqnarray}
Here, $E_0$ and $E_0^s$ 
are the single particle ground-state energies of the non-strange and 
strange quark, respectively. 
These mass formulas satisfy the Gell-Mann-Okubo mass relations  
\begin{eqnarray}\label{GMO} 
&&m_{\Sigma} - m_N = \frac12 (m_{\Xi}-m_N) + 
 \frac34 (m_{\Sigma}-m_{\Lambda})\,,  \hspace*{.25cm} 
m_{\Sigma^\ast} - m_{\Delta} = m_{\Xi^\ast}-m_{\Sigma^\ast} = 
m_{\Omega} - m_{\Xi^\ast}
\end{eqnarray}
providing a condition on the matrix elements of the residual 
interaction with  
${\cal H}_{qq} - {\cal H}_{qs} \simeq {\cal H}_{qs} - {\cal H}_{ss}$.
With the choice $E_0^s - E_0 \simeq 180$ MeV, ${\cal H}_{qq} \simeq 400$ MeV and 
${\cal H}_{qq} - {\cal H}_{qs} \simeq {\cal H}_{qs} - {\cal H}_{ss} \simeq 150$ MeV,
all the observed mass differences can be roughly reproduced.

The hyperfine interaction of Eq.~(\ref{eqn:hyperint}) can for example 
be generated by the one-gluon-exchange mechanism between two quarks 
(in the non-relativistic reduction)~\cite{DeRujula:ge}. Actually, 
the relevant part for the ground-state baryons is just given by the 
form of Eq.~(\ref{eqn:hyperint}), and 
the mass differences in the multiplet have been  
explained~\cite{DeRujula:ge}. Moreover, mass and structure of excited  
p-wave baryons have been studied successfully by taking into account 
the hyperfine interaction including a tensor term involved in the 
one-gluon-exchange~\cite{Isgur:xj}.

During the last decade quark models based on one-meson-exchange 
responsible for the inter-quark forces are often used as an
alternative to study 
the baryon mass spectrum 
and various consequences resulting from the structure of these
states~\cite{Glozman:1995fu,Glozman:1997ag,Glozman:1997fs,Glozman:fu,Valcarce:1995dm,Fernandez:kz,Garcilazo:md,Garcilazo:2003wq,Furuichi:2002gi}. 
The relevant short range part of the force is essentially the hyperfine 
interaction of the form 
$\vec \lambda_i \cdot \vec \lambda_j \, \vec s_i 
\cdot \vec s_j \,\delta^3(\vec r_i -\vec r_j)$, where an 
explicit flavor dependence is implemented by the flavor operators 
$\vec \lambda$, and hence in its flavor dependence is different 
from Eq.~(\ref{eqn:hyperint}). In these works the baryon spectrum 
including the splitting within flavor multiplets, is quantitatively 
explained by this force instead of Eq.~(\ref{eqn:hyperint}), when 
adjusting the radial dependence of the interaction phenomenologically.
Both of these two mechanisms for explaining the flavor and spin 
dependence of baryon masses are based on the 
non-relativistic picture with massive constituent quarks. 

The importance of chiral symmetry for the phenomenology of hadron physics
at low and intermediate energies has been worked out and strongly 
established. The spontaneous breakdown of chiral symmetry leads to the 
presence of Goldstone bosons, which are realized by the low-lying octet 
of pseudoscalar mesons. Therefore, it is natural to expect that the 
Goldstone bosons play a significant role in the description of the 
spectrum of baryons and their structure. However, other mechanisms, 
such as one-gluon exchange, still can play some role.

The Graz group~\cite{Glozman:1997fs} claims that a possible 
gluon-exchange is excluded in their quark model which is solely based 
on Goldstone dynamics, while the Salamanca-Valencia 
approach~\cite{Valcarce:1995dm} resorts to a strong residual gluon 
interaction. In both works the key justification for the respective 
flavor- and spin-dependent quark interactions rests on the correct 
description of the level ordering in the nucleon spectrum, \ie the 
positive parity Roper resonance (1440) moves below the negative parity
$N^\ast$ (1535) by the effect of the residual interaction.
Actually, results and conclusions drawn strongly 
depend on the treatment of the kinetic term (\ie non-, semi-, or 
full-relativistic) and of the short-range interaction
(contact term or smeared out interaction)~\cite{Garcilazo:2003wq}. 
In addition, in the quark model approach it is assumed that resonances 
are exclusively described as simple excited three quark systems.

Dynamical generation of resonances in multichannel approaches are an 
alternative picture to possibly explain the excitation spectrum of 
baryons. For example, in the chiral unitary approach, the $N^\ast(1535)$ 
resonance is mainly a $K \Sigma$ and $K \Lambda$ 
system~\cite{Kaiser:1995cy,Nacher:1999vg,Nieves:2001wt,Inoue:2001ip}. 
This approach suggests that the coupling to configurations by creating 
a $s\bar s$ pair is essential to explain the resonance structure in the 
quark picture. Also, an even more complicated multi-channel structure 
is seemingly underlying the Roper resonance~\cite{Capstick:xn,Cardarelli:1996vn,Krehl:1999km,Hernandez:2002xk}.

In this paper we study the mass spectrum of the so-called ground-state 
baryons in a relativistic chiral quark model. Here we restrict to the
octet and decuplet ground states of flavor SU(3) baryons, since
the valence quark component is the leading contribution and the
residual interaction can be treated perturbatively.

The perturbative chiral quark model~\cite{Lyubovitskij:2001nm,Lyubovitskij:2000sf,Lyubovitskij:2001fv,Inoue:2003bk}
is an effective model of baryons based on chiral symmetry.
The baryon is described as a state of three localized relativistic 
quarks supplemented by a pseudoscalar meson cloud as dictated by chiral 
symmetry requirements. In this model the effect of the meson cloud is 
evaluated perturbatively in a systematic fashion.
The model has been successfully applied to the nucleon 
electro-magnetic form factors, the meson-nucleon sigma term, the 
nucleon-$\Delta$ photo-transition and the nucleon axial coupling 
constant amongst other applications. 
In this paper, we extend the model to include gluon degrees 
of freedom and we also introduce a violation of flavor $SU(3)$ symmetry. 
Then we evaluate meson and gluon cloud induced mass shifts for the 
octet and decuplet of ground-state baryons. 

This paper is organized as follows. In Section~\ref{Sect_PCQM}, 
we introduce the perturbative chiral quark model with additional gluon 
degrees of freedom and the flavor violation.
In Section~\ref{Sect_Shifts}, we study the masses of the 
ground-state baryon within the model. 
Section~\ref{Sect_summary} contains a summary of our major 
conclusions. 

\section{The perturbative chiral quark model (PCQM)}
\label{Sect_PCQM} 
The perturbative chiral quark model~\cite{Lyubovitskij:2001nm,Lyubovitskij:2000sf,Lyubovitskij:2001fv,Inoue:2003bk}
is based on an effective chiral Lagrangian describing the 
valence quarks of baryons as relativistic fermions moving in an 
external field (static potential) $V_{\rm eff}(r)=S(r)+\gamma^0 V(r)$
with $r=|\vec x|$~\cite{Lyubovitskij:2001nm,Lyubovitskij:2000sf},  
which in the $SU(3)$-flavor version are supplemented by a cloud of 
Goldstone bosons $(\pi, K, \eta)$. The origin of a such an effective 
potential can be phenomenologically understood by the assumptions 
of Refs.~\cite{Leutwyler:1980ma} where the gluon field can be 
decomposed into the vacuum background configuration and small 
quantum fluctuations around it. In particular, the vacuum component 
of the gluon field provides the confinement of quarks, which in turn is
encoded in a confinement potential. The quantum fluctuations of the 
gluon field can be also taken into account by dressing the quark fields 
using a perturbative expansion.  
 
When treating Goldstone fields and the quantum components of the gluon 
fields as small fluctuations around the three-quark (3q) core, we have 
the linearized effective 
Lagrangian~\cite{Lyubovitskij:2001nm,Lyubovitskij:2000sf}: 
\eq\label{linearized_L}
{\cal L}_{\rm eff}(x) &=&
\bar{\psi}(x) [i \not\!\partial - V_{\rm eff}(r)]\psi(x) +
\frac{1}{2} \sum\limits_{i=1}^{8} [\partial_\mu \Phi_i(x)]^2 
- \frac{1}{4} F^a_{\mu\nu} F^{a \, {\mu\nu}} \nonumber\\ 
 &-& \bar{\psi}(x) \biggl\{ 
S(r) i \gamma^5 \frac{\hat \Phi (x)}{F} + 
g_s \gamma^\mu A_\mu^a(x) \frac{\lambda^a}{2} \biggr\}
 \psi(x)+{\cal L}_{SB}(x).
\en 
$F=88$ MeV is the pion decay constant in the chiral 
limit~\cite{Gasser:1987rb}; $g_s$ is the quark-gluon coupling constant;  
$A_\mu^a$ is the quantum component of the gluon field and 
$F^a_{\mu\nu}$ is its conventional field strength tensor;   
$\hat\Phi = \sum\limits_{i=1}^{8} \Phi_i \lambda_i = \sum\limits_P 
\Phi_P \lambda_P$ is the octet matrix of pseudoscalar mesons with 
$P = \pi^\pm, \pi^0, K^\pm, K^0, \bar K^0, \eta$. The explicit relations 
between the sets $\{\Phi_P, \lambda_P\}$ and $\{\Phi_i, \lambda_i\}$  
are given in Ref.~\cite{Inoue:2003bk}. 

The term ${\cal L}_{SB} = {\cal L}_{\chi SB} \, + \, 
\bar{\cal L}_{SB}$ in Eq.~(\ref{linearized_L}) contains the mass 
contributions ${\cal L}_{\chi SB}$ both for quarks and mesons, 
which explicitly break chiral symmetry, 
\begin{equation}
{\cal L}_{\chi SB}(x) = - \bar\psi(x) {\cal M} \psi(x)
- \frac{B}{2} Tr [\hat \Phi^2(x)  {\cal M} ]\,,
\end{equation} 
and the term $\bar{\cal L}_{SB}$, which breaks 
the unitary flavor $SU(3)$ symmetry: 
\begin{equation}
\bar{\cal L}_{SB}(x) = 
- \bar\psi(x) \Delta V_{\rm eff}(r) \lambda_s \psi(x)\,.
\end{equation} 
Here, ${\cal M}={\rm diag}\{m_u,m_d,m_s\}$
is the mass matrix of current quarks, 
$B=-\langle 0|\bar u u|0 \rangle / F^2$ 
is the quark condensate constant, $\Delta V_{\rm eff}(r) = 
\Delta S(r) + \gamma^0 \Delta V(r)$ is a correction to the 
confinement potential for strange quarks and 
$\lambda_s = {\rm diag}\{0, 0, 1\}$ is the strangeness matrix. 
We suppose that the nonstrange and the strange quarks are exposed to 
a different confinement and, therefore, introduce flavor dependent 
phenomenological potentials: $V_{\rm eff}(r)$ for nonstrange quarks and 
$V_{\rm eff}(r) + \Delta V_{\rm eff}(r)$ for the strange quark. 
In practice, it is convenient to separate the term 
$$\bar s \Delta V_{\rm eff}(r) s = 
\bar\psi \Delta V_{\rm eff}(r) \lambda_s \psi$$  
and treat it as an operator which breaks the flavor $SU(3)$ symmetry. 
After diagonalization the meson mass 
terms take the form 
\begin{equation}
\frac{B}{2} \, {\rm Tr} [\hat \Phi^2(x)  {\cal M} ] = 
\frac{1}{2} \sum_P \, M_P^2 \, \Phi_P^2(x) 
\end{equation} 
where $M_P$ is the set of pseudoscalar meson masses. 
In the numerical calculations we restrict to the isospin symmetry 
limit with $m_u=m_d=\hat m$. 
We rely on the standard picture of chiral symmetry 
breaking~\cite{Gasser:1982ap} and for the masses of pseudoscalar 
mesons we use the leading term in their chiral expansion, i.e. 
linear in the current quark masses (see exact expressions in 
Ref.~\cite{Inoue:2003bk}. In the isospin limit they are given by 
\eq\label{M_Masses}
M_{\pi}^2=2 \hat m B, \hspace*{.5cm} M_{K}^2=(\hat m + m_s) B,
\hspace*{.5cm} M_{\eta}^2= \frac{2}{3} (\hat m + 2m_s) B.
\en 
The following set of
parameters~\cite{Gasser:1982ap} is chosen in our evaluation
\begin{equation}
\hat m = 7 \;{\rm MeV},\; \frac{m_s}{\hat m}=25,\;
B = \frac{M^2_{\pi^+}}{2 \hat m}=1.4 \;{\rm GeV}.
\end{equation}
Meson masses obtained by Eq.~(\ref{M_Masses}) satisfy the
Gell-Mann-Oakes-Renner and the Gell-Mann-Okubo relation. In addition,
the linearized effective Lagrangian in Eq.~(\ref{linearized_L})
fulfills the PCAC requirement.

We expand the quark field $\psi$ in the basis of potential
eigenstates as
\eq\label{total_psi}
\psi(x) = \sum\limits_\alpha b_\alpha u_\alpha(\vec{x})
\exp(-i{\cal E}_\alpha t) + \sum\limits_\beta
d_\beta^{\dagger} v_\beta(\vec{x}) 
\exp(i{\cal E}_\beta t)\, ,
\en
where the sets of quark $\{ u_\alpha \}$ and antiquark $\{ v_\beta \}$
wave functions in orbits $\alpha$ and $\beta$ are solutions of the
Dirac equation with the static potential 
$V_{\rm eff}(r) + \Delta V_{\rm eff}(r) \lambda_s$. 
The expansion coefficients $b_\alpha$ and 
$d_\beta^{\dagger}$ are the corresponding single quark annihilation 
and antiquark creation operators.

We formulate perturbation theory in the expansion parameter 
$\hat\Phi(x)/F \sim 1/\sqrt{N_c}$ and treat finite current quark 
masses perturbatively~\cite{Lyubovitskij:2001nm}. 
All calculations are performed at one loop or at order of accuracy 
$o(1/F^2, \hat{m}, m_s)$. In the calculation of matrix elements we 
project quark diagrams on the respective baryon states. The baryon 
states are conventionally set up by the product of the $SU(6)$ 
spin-flavor and $SU(3)_c$ color wave functions. Then the 
nonrelativistic single 
quark spin wave function is replaced by the relativistic solution
$u_\alpha(\vec{x})$ of the Dirac equation 
\begin{eqnarray}\label{Dirac_eq_q}
\left[ -i\gamma^0\vec{\gamma}\cdot\vec{\nabla} + \gamma^0 S(r) + V(r)
- {\cal E}_\alpha^q \right] u_\alpha^q(\vec{x})=0 
\end{eqnarray}
for nonstrange quarks, and 
\begin{eqnarray}\label{Dirac_eq_s}
\left[ -i\gamma^0\vec{\gamma}\cdot\vec{\nabla} 
+ \gamma^0 \{ S(r) + \Delta S(r) \} + V(r) + \Delta V(r) 
- {\cal E}_\alpha^s \right] u_\alpha^s(\vec{x})=0
\end{eqnarray} 
for the strange quark. 
Here, index $q$ refers to the nonstrange ($u$ or $d$) whereas $s$ to
the strange quark; $\{u_\alpha^q, {\cal E}_\alpha^q\}$ and 
$\{u_\alpha^s, {\cal E}_\alpha^s\}$ are the single-quark wave functions 
and energies of nonstrange and strange quark, respectively.  

For the description of baryon properties on tree level we use the 
effective potentials $V_{\rm eff}(r)$ 
and $\Delta V_{\rm eff}(r)$ with a quadratic radial 
dependence~\cite{Lyubovitskij:2001nm,Lyubovitskij:2000sf}: 
\eq\label{V_eff}
S(r) &=& M_1 + c_1 r^2, \hspace*{2cm} V(r) = M_2+ c_2 r^2 \\
\Delta S(r) &=& \Delta M_1 + \Delta c_1 r^2, \hspace*{1cm} 
\Delta V(r) = \Delta M_2+ \Delta c_2 r^2 
\en 
with the particular choice
\eq
&&M_1 = \frac{1 \, - \, 3\rho^2_q}{2 \, \rho_q R_q} , \hspace*{1.7cm}
M_2 = {\cal E}_0^q - \frac{1 \, + \, 3\rho^2_q}{2 \, \rho_q \, R_q}\,, 
\hspace*{1.7cm}
c_1 \equiv c_2 =  \frac{\rho_q}{2R^3_q}\,, \\
&&M_1 + \Delta M_1 = \frac{1 \, - \, 3\rho^2_s}{2 \, \rho_s \, R_s}\,, 
\hspace*{.3cm} 
M_2 + \Delta M_2 = {\cal E}_0^s - 
\frac{1 \, + \, 3\rho^2_s}{2 \, \rho_s \, R_s} , \hspace*{.3cm} 
c_1 + \Delta c_1 \equiv c_2 + \Delta c_2 = \frac{\rho_s}{2R^3_s} \,.   
\en
Here, the potential parameters are related to the set of quantities
$(R_q, \rho_q)$ and $(R_s, \rho_s)$, which characterize the
ground-state wave function of nonstrange $(u_0^q)$ 
and strange quark $(u_0^s)$: 
\eq\label{Gaussian_Ansatz}
u_0^q(\vec{x}) \, &=& 
\, N_q \, \exp\biggl[-\frac{\vec{x}^{\, 2}}{2R^2_q}\biggr]
\, \left(
\begin{array}{c}
1\\
i \rho_q \, \vec{\sigma}\vec{x}/R_q\\
\end{array}
\right)
\, \chi_s \, \chi_f \, \chi_c\,, \\
u_0^s(\vec{x}) \, &=& \, N_s \, 
\exp\biggl[-\frac{\vec{x}^{\, 2}}{2R^2_s}\biggr]
\, \left(
\begin{array}{c}
1\\
i \rho_s \, \vec{\sigma}\vec{x}/R_s\\
\end{array}
\right)
\, \chi_s \, \chi_f \, \chi_c, 
\label{Gaussian_Ansatz_s}
\en
where 
$N_q=[\pi^{3/2} R^3_q (1+3\rho^2_q/2)]^{-1/2}$ and 
$N_s=[\pi^{3/2} R^3_s (1+3\rho^2_s/2)]^{-1/2}$ are the 
corresponding normalization constants; $\chi_s$, $\chi_f$, $\chi_c$ 
are the spin, flavor and color quark wave functions, respectively. 
The constant parts of the scalar potential $M_1$ and $M_1 + \Delta M_1$ 
can be interpreted as the constituent masses of the quarks,
which are simply the displacements of 
the current quark masses $\hat m$ and $m_s$ due to the potential 
$S(r) + \Delta S(r) \lambda_s$. 

In previous works, for example in the calculation of electromagnetic
nucleon form factors~\cite{Lyubovitskij:2001nm}, we restricted our 
kinematics to a specific reference frame, that is the Breit frame.
This particular choice is sufficient
to guarantee local gauge invariance concerning the 
coupling of the electromagnetic field (for a detailed discussion see 
Ref.~\cite{Lyubovitskij:2001nm}). The Breit frame is specified as 
follows: the momentum of the initial state is $p = (E, -\vec{q}/2)$, 
the final momentum is $p^\prime = (E, \vec{q}/2)$ and the 4-momentum of 
the external field is $q = (0, \vec{q}\,)$ with $p^\prime = p + q$.
The constraint of local gauge invariance requires that the 
bare energies of the nonstrange $({\cal E}_0^q)$ 
and strange $({\cal E}_0^s)$ quarks should be equal to each other: 
${\cal E}_0^q = {\cal E}_0^s = {\cal E}_0$.  
In the following we use this requirement for consistency of our 
calculation. Unitary symmetry breaking at tree 
level is then contained in the deformation of the bare wave function of 
the strange quark with respect to the nonstrange one: 
$u_0^s(\vec{x}) \not= u_0^q(\vec{x})$.  

The parameters $\{\rho_q, R_q\}$ and $\{\rho_s, R_s\}$ can be fixed
from a study of some canonical (electromagnetic and semileptonic)  
properties of baryons at zeroth order. 
The parameter $\rho_q$ is related to the nucleon axial charge $g_A$ 
calculated in zeroth-order (or 3q-core) approximation:
\eq\label{ga_rho_match}
g_A=\frac{5}{3}\biggl(1 - 
\frac{2\rho^2_q}{1+\frac{3}{2}\rho^2_q}\biggr) 
   =\frac{5}{3}\biggl(1 - \frac{2}{3}[1 - \gamma_q] \biggr) 
\en 
where 
\eq\label{gamma}
\gamma_i = \frac{1 - \frac{3}{2} \rho^2_i}{1 + \frac{3}{2} \rho^2_i}
\en 
is the relativistic reduction factor for nonstrange $(i=q)$ 
and strange $(i=s)$ quarks (the specific value $\gamma_i = 1$ 
corresponds to the nonrelativistic limit)~\cite{Donoghue:dd}. 
Therefore, $\rho_q$ can be replaced by $g_A$ using the matching 
condition (\ref{ga_rho_match}). In our calculations we use the value 
$g_A$=1.25~\cite{Lyubovitskij:2001nm,Lyubovitskij:2000sf} or 
$\rho_q = \sqrt{2/13} \simeq 0.392$.  
The parameter $R_q$ is related to the charge radius
of the proton in the zeroth-order approximation as
\eq\label{rad_LO}
\langle r^2_E \rangle^P_{LO} = \int d^3 x \,u^{q\, \dagger}_0(\vec{x})\,
\vec{x}^{\, 2}\,u^q_0(\vec{x})\,=\,\frac{3R^2_q}{2}\,
\frac{1\,+\,\frac{5}{2}\,\rho^2_q}{1\,+\,\frac{3}{2}\,\rho^2_q}.
\en
In previous numerical studies $R_q$ is varied in the region
from 0.55 fm to 0.65 fm, which 
corresponds to a change of $\langle r^2_E \rangle^P_{LO}$ from 
0.5 to 0.7 fm$^2$~\cite{Lyubovitskij:2001nm,Lyubovitskij:2000sf}.  
In the current work we use the central value of $R_q=0.6$ fm.

The values of the parameters $\rho_s$, $R_s$ can be deduced from the 
ratios $g_A/g_V$ of the axial $(g_A)$ and vector $(g_V)$
constants in semileptonic decays of hyperons.
In zeroth-order approximation following relations 
for the $g_A/g_V$ ratios are obtained: 
\eq 
& &(g_A/g_V)|_{\Lambda \to p e^- \bar\nu_e} = - r_A 
\,\,\,\,\,\,\,\,\,\,\,\,\,\,\,\, 
   (- 0.718 \pm 0.015)\,,\nonumber\\ 
& &(g_A/g_V)|_{\Sigma^- \to n e^- \bar\nu_e} = \frac{1}{3} \, 
r_A \,\,\,\,\,\,\,\,\,\,\,\,\, (0.32 \pm 0.017)\,,\nonumber\\ 
& &(g_A/g_V)|_{\Xi^- \to \Lambda^0 e^- \bar\nu_e} = 
- \frac{1}{3} \, r_A \,\,\,\,\,\, (- 0.25 \pm 0.05)\,,\nonumber\\ 
& &(g_A/g_V)|_{\Xi^0 \to \Sigma^+ e^- \bar\nu_e} = - \frac{5}{3} \, 
r_A \,\,\,\,\,\, (- 1.32^{+0.21}_{-0.17} \pm 0.05)\,.\nonumber 
\en 
In the brackets $(\ldots)$ we quote the correponding experimental 
value~\cite{Hagiwara:fs}. The quantity $r_A$ is the ratio of the spatial 
matrix elements of the axial and vector quark currents: 
\eq
r_A \, = \, 
\frac{\displaystyle{1 \, - \, \frac{\rho^2_q}{2} \, 
\kappa_s}}{\displaystyle{1 \, + \, \frac{3}{2} \, 
\rho^2_q \, \kappa_s}} = 1 \, - \, \frac{2}{3} \, (1 - \gamma_{qs})\,, 
\,\,\,\,\,\,\,\,\,\,\,
\kappa_s \, = \, 
\frac{2\Delta_s}{1 \, + \, \Delta^2_s} \, \frac{\rho_s}{\rho_q} \,, 
\,\,\,\,\,\,\,\,\,\,\, \Delta_s \, = \, \frac{R_s}{R_q}   
\en 
which is related to the relativistic reduction factor $\gamma_s$ 
for the $s \to u$ flavor exchange: 
\eq 
\gamma_{qs} \, = \, 
\frac{\displaystyle{1 \, - \, \frac{3}{2} \, \rho^2_q \,\kappa_s}} 
{\displaystyle{1\,+\,\frac{3}{2}\, \rho^2_q \,\kappa_s}} 
\en 
In the $SU(3)$ flavor limit with $R_s = R_q$ and 
$\rho_s = \rho_q$  (or $\kappa_s = 1$) the factor $\gamma_{qs}$ 
reduces to $\gamma_q$ introduced in Eq.~(\ref{gamma}) 
for the $d \to u$ flavor transition. 
In this study, the parameters $\rho_s$ and $R_s$ are fitted, as will be 
shown later,  by using the baryon mass spectrum as well as to the data 
for the $g_A/g_V$ ratios. 
The best choice is: $\rho_s = 0.354 $ and $R_s = 0.58 $ fm. 
This results in the value of $r_A = 0.77$, which satisfies the above 
constraint from semileptonic hyperon decay.

To evaluate the mass shifts of the baryons due to the residual 
interaction we formulate perturbation theory.
In general, the expectation value of an operator $\hat A$ is set up as:
\begin{equation}\label{perturb_A}
\langle \hat A \rangle = {}^B \langle \phi_0 |\sum^{\infty}_{n=0} 
\frac{i^n}{n!}\int d^4 x_1 \ldots \int d^4 x_n T[{\cal L}_I (x_1) 
\ldots {\cal L}_I (x_n) \hat A]|\phi_0 \rangle^B_c,
\end{equation}
where the state vector $|\phi_0 \rangle$ corresponds to the unperturbed
three-quark state ($3q$-core). Superscript $"B"$ in~(\ref{perturb_A}) 
indicates that the matrix elements have to be projected onto the 
respective baryon states, whereas subscript $"c"$ refers to 
contributions from connected graphs only. ${\cal L}_I (x)$ contained 
in Eq.~(\ref{perturb_A}) is the interaction Lagrangian derived 
in Eq.~(\ref{linearized_L}):
\begin{eqnarray}
{\cal L}_I (x) =  - \bar{\psi}(x) \biggl\{ 
S(r) i \gamma^5 \frac{\hat \Phi (x)}{F} + 
g_s \gamma^\mu A_\mu^a(x) \frac{\lambda^a}{2} \biggr\} \,.
\end{eqnarray}
For the evaluation of Eq.(\ref{perturb_A}) we apply Wick's
theorem with the appropriate propagators for quarks and mesons.

For the quark field we use a Feynman
propagator for a fermion in a binding potential with
\eq\label{quark_propagator}
\delta^{mn} i G_\psi^m(x,y) = 
\langle 0|T\{\psi^m(x) \bar\psi^n(y)\}|0 \rangle
\en 
where $m, n$ are the flavor indices and 
\eq
i G_\psi^m(x,y) &=& \theta(x_0-y_0) \, \sum\limits_{\alpha} \, 
u_\alpha^m(\vec{x}) \, \bar u_\alpha^m(\vec{y}) \, 
e^{-i{\cal E}_\alpha^m (x_0-y_0)}\nonumber\\
&-& \theta(y_0-x_0) \, \sum\limits_{\beta} \, v_\beta^m(\vec{x}) \, 
\bar v_\beta^m(\vec{y}) \, e^{i{\cal E}_\beta^m (x_0-y_0)} \,.
\en
In present study we restrict
the expansion of the quark propagator to the ground state with:
\eq\label{quark_propagator_ground}
i G_\psi^m(x,y) \to i G_0^m(x,y) \doteq u_0^m(\vec{x}) \, 
\bar u_0^m(\vec{y}) \, e^{-i{\cal E}_0^m (x_0-y_0)} \, 
\theta(x_0-y_0).
\en
Such a truncation can be considered as an additional regularization of 
the quark propagator, where in the case of flavor $SU(2)$ intermediate 
baryon states in loop-diagrams are restricted to $N$ and $\Delta$. 

For the meson fields we adopt the free Feynman propagator with
\eq
i\Delta_{PP^\prime}(x-y) = 
\langle 0|T\{\Phi_P(x)\Phi_{P^\prime}(y)\}|0 \rangle = 
\delta_{PP^\prime}\int\frac{d^4k}{(2\pi)^4i}
\frac{\exp[-ik(x-y)]}{M_P^2 - k^2 - i\epsilon}. 
\en

For the gluon field we use the dressed propagator containing an 
effective quark-gluon coupling constant 
$\alpha_s(k^2) = g_s^2(k^2)/(4\pi)$ 
with a nontrivial momentum dependence. In our consideration we use the 
Coulomb gauge\footnote{It can be shown that the results do not depend 
on the choice of the gauge.} to separate the contributions of Coulomb 
($A_0^a$) and transverse ($A_i^a$) gluons in the propagator: 
\begin{eqnarray}\label{D00}
\hspace*{-.75cm}
i \, \alpha_s \, D_{00}^{ab}(x-y) \, = \,   
- \, \delta^{ab} \int\frac{d^4k}{(2\pi)^4i} 
\frac{e^{-ik(x-y)}}{\vec{k}^{\, 2}} \, 
\alpha_s(k^2) 
\end{eqnarray}
and 
\begin{eqnarray}\label{Dij}
\hspace*{-.75cm}
i \, \alpha_s \, D_{ij}^{ab}(x-y) \, = \,  
\delta^{ab} \int\frac{d^4k}{(2\pi)^4i} 
\frac{e^{-ik(x-y)}}{k^2 + i \varepsilon} 
\biggl\{ \delta_{ij} - \frac{k_ik_j}{\vec{k}^{\, 2}} \biggr\} \, 
\alpha_s(k^2) \,.   
\end{eqnarray} 
Following the ideas of the approach to low-energy QCD based on 
the solutions of the Dyson-Schwinger equations~\cite{Maris:1999nt} 
we suppose that the running coupling constant $\alpha_s(k^2)$ 
includes nontrivial effects of vertex and self-energy corrections, etc. 
In the present paper, we test a simple analytic form for $\alpha_s$ 
as suggested in Ref.~\cite{Maris:1999nt}:  
\begin{eqnarray}
\alpha_s(t) \, = \, \frac{\pi}{\omega^6} D {t}^2 
 e^{- t/\omega^2} + \frac{2 \gamma_m \pi}
{\ln\biggl[\tau + \left(1 + t/\Lambda_{QCD}^2\right)^2 
\biggr]} F(t)
\end{eqnarray}
where $t = -k^2$ is an Euclidean momentum squared,  
$F(t)= 1-\exp(-t/[4m_t^2])$\,, $\tau = e^2-1$\,, 
$\gamma_m = 12/(33 - 2 N_f)$\,,  $\Lambda_{QCD}^{N_f=4}=0.234$ GeV\,, 
$\omega = 0.3$ GeV\, and $m_t = 0.5$ GeV.  
The functional form of $\alpha_s(t)$ was fitted to the perturbative 
QCD result in the ultraviolet region (at large momentum squared), 
and is governed by a single parameter $D$ in the infrared region. 
In Ref.~\cite{Maris:1999nt} the parameter $D = (0.884\,\mbox{GeV})^2$ 
was adjusted phenomenologically to reproduce properties of pions and 
kaons described as bound states of constituent quarks. In our 
considerations we fit the effective coupling with 
$D = (0.49\,\mbox{GeV})^2$ such that the $\Delta-N$ mass splitting is 
reproduced. The running coupling $\alpha_s(t)$ between quarks and 
gluons should be considered as effective, since it depends on the way 
its used in phenomenology. Since, in our model, we utilize gluon as 
well as meson degrees of freedom,
it is natural that our fitted value for $D$ is smaller than the one of
Ref.~\cite{Maris:1999nt}. Namely, the relatively small value for $D$ 
leaves space for meson cloud effects in the infrared region.

\section{Baryon mass shift with in the PCQM}
\label{Sect_Shifts} 
In chiral quark models, the mass of the ground-state baryon 
is given by sum of the contribution of the three-quark core, 
the finite current quark mass and the mass shift
due to the interaction between quarks and meson fields. 
In the present paper, we also include the gluon induced 
mass shifts, hence the ground-state baryon mass is written as
\begin{equation}
m_{B} \, = \, E_{3q}\, + \, \sum\limits_i \gamma_i m_i 
\, + \, \Delta m_B^{M} \, + \, \Delta m_B^{G} ~.
\end{equation} 
The first term $E_{3q}$ is the mass of the three-quark 
core which is the sum of the bare energies of the ground-state 
quarks including the subtraction of the spurious center-of-mass (cm) 
term $E_{\rm cm}$: 
\begin{equation}\label{E_3q}  
E_{3q} = 3 {\cal E}_0 - E_{\rm cm}\,. 
\end{equation}   
The second term is the finite current quark mass contribution, 
the last terms are the meson and gluon cloud induced mass 
shifts, respectively. 

The bare energy of the ground-state quark is the same in magnitude 
for both the nonstrange and strange quarks by construction 
(see discussion in Section~\ref{Sect_PCQM}) and is considered as 
a free parameter. We also use a unified value of $E_{\rm cm}$ 
for all baryons. Since ${\cal E}_0$ and  $E_{\rm cm}$ appear
in a linear combination in Eq.~(\ref{E_3q}), we only have
one free parameter - the mass of the three-quark core $E_{3q}$.   
We fix $E_{3q}$ from a fit of the nucleon mass by including
all one-loop effects: $E_{3q} \, = \, 1612.4 \,\, {\rm MeV} $.
The shift of the quark energy due to the finite current quark
mass $m_i = \hat{m}\,, \, m_s$ is calculated perturbatively
and is given by a term linear in $m_i$ 
(see details in Refs.~\cite{Lyubovitskij:2001nm,Lyubovitskij:2000sf}): 
\eq 
{\cal E}_0 \, \to \, {\cal E}_0(m_i) \, = \, 
{\cal E}_0 \, + \, \gamma_i \, m_i + o(m_i)\,.  
\en
where $m_i$ and $\gamma_i$ are the
values for the current quark mass ($\hat m$ or $m_s$) and for the 
relativistic reduction factor ($\gamma_q$ or $\gamma_s$). From the exact 
solution of the Dirac equation including the current quark 
mass term it can be shown that higher order corrections $o(m_i)$ are 
negligible. In the case of the nonstrange quark the linear term 
$\gamma_q \hat{m}$ gives a correction of the order of $1\%$, whereas
the term with $o(\hat{m})$ is about $10^{-2} \%$ of ${\cal E}_0$. For 
the strange quark the linear term is more important yielding a 
correction of $\sim 20\%$, the higher-order term $o(m_s)$ is suppressed 
giving a contribution of only $2\%$ of ${\cal E}_0$. 

In the PCQM, the mass shift 
$\Delta m_B = \Delta m_B^{M} \, + \, \Delta m_B^{G}$  
is evaluated perturbatively. In the one-loop approximation it is 
given by 
\begin{eqnarray}
 \Delta m_B  =
 {}^B\!\langle \phi_0| \sum_{n=1}^{2} \frac{i^n}{n!} 
  \! \int \! i \delta(t_1) d^4 x_1 \cdots d^4 x_n  
 T[{\cal L}_I (x_1) \cdots {\cal L}_I(x_n) ]
 | \phi_0 \rangle^B_c 
\label{eqn:gafull}
\end{eqnarray}
where ${\cal L}_I$ is the interaction Lagrangian of quarks 
with meson and gluon fields. The diagrams contributing to
$\Delta m_B$ are shown in Fig.1 (meson contribution) and 
in Fig.2 (gluon contribution).    

The meson induced  baryon mass shift $\Delta m_B^M$ 
consists of the contribution of the meson cloud (Fig.1a) and
the meson exchange (Fig.1b) diagram. 
The expression for the meson contribution to the baryon mass 
shift is given by 
\eq
\Delta m_B^M \, = \, \!\!\! \sum_{\Phi=\pi,K,\eta} \,  
\Delta m_B^{\Phi} \, = \, \!\!\! \sum_{\Phi=\pi,K,\eta} \,  
[ \, \Delta m_B^{\Phi {\rm [C]}}(M_\Phi^2) \, + \, 
\Delta m_B^{\Phi {\rm [E]}}(M_\Phi^2) \, ]   
\label{eqn:pcqmmassshifts}
\en 
where $\Delta m_B^{\Phi {\rm [C]}}$ is the contribution of the 
diagram of Fig.1a and $\Delta m_B^{\Phi {\rm [E]}}$ is related to
the diagram of Fig.1b. The partial contribution  
of the $\pi$, $K$, and $\eta$-meson cloud to the baryon mass shifts 
are written as
\eq 
& &    \Delta m_B^{\pi {\rm [C]}}(M_\pi^2) = 
 d_{B}^{\pi {\rm [C]}}  
\, \Pi_{[qq, qq]}(M_\pi^2)\\
&&\nonumber\\
& &    \Delta m_B^{K {\rm [C]}}(M_K^2) = 
 d_{B}^{K {\rm [C1]}}  
\, \Pi_{[qs, sq]}(M_K^2) \, + \, 
 d_{B [sq, qs]}^{K {\rm [C2]}}  
\, \Pi_{[sq, qs]}(M_K^2)\\
&&\nonumber\\
& &    \Delta m_B^{\eta {\rm [C]}}(M_\eta^2) = 
 d_{B}^{\eta {\rm [C1]}}  
\, \Pi_{[qq, qq]}(M_\eta^2) \, + \, 
 d_{B}^{\eta {\rm [C2]}}  
\, \Pi_{[ss, ss]}(M_\eta^2)\\
&&\nonumber\\
& &    \Delta m_B^{\pi {\rm [E]}}(M_\pi^2) = 
 d_{B}^{\pi {\rm [E]}}  
\, \Pi_{[qq, qq]}(M_\pi^2)\\
&&\nonumber\\
& &    \Delta m_B^{K {\rm [E]}}(M_K^2) = 
 d_{B}^{K {\rm [E1]}} \, 
\Pi_{[qs, sq]}(M_K^2) \, + \, 
 d_{B}^{K {\rm [E2]}} \, 
\Pi_{[sq, qs]}(M_K^2)\\
&&\nonumber\\
& &    \Delta m_B^{\eta {\rm [E]}}(M_\eta^2) =  
 d_{B}^{\eta {\rm [E1]}}  
\, \Pi_{[qq, qq]}(M_\eta^2) \, + \, 
 d_{B}^{\eta {\rm [E2]}}  
\, \Pi_{[qq, ss]}(M_\eta^2) \, + \, 
 d_{B}^{\eta {\rm [E3]}}  
\, \Pi_{[ss, ss]}(M_\eta^2) . 
\en 
Here, $d_{B}^{\Phi {\rm [J]}}$ are recoupling coefficients 
which are listed in Table~\ref{tbl:dcoeff} and 
$\Pi_{[ij, kl]}(M_\Phi^2)$ are the meson-loop integrals 
\eq 
\Pi_{[ij, kl]}(M_\Phi^2) \, = \, 
- \biggl(\frac{g_A}{\pi F}\biggr)^2 \,\,\int\limits_0^\infty dp\, p^4\,  
\frac{G_{ij}(p^2) \, G_{kl}(p^2)}{p^2 \, + \, M_\Phi^2} \,. 
\en
The quark-meson transition form factor $G_{ij}(p^2)$ is given by  
\eq 
G_{ij}(p^2)\,=\,\rho_{ij}\,\biggl(\frac{\Delta_{ij}} 
{\Delta_{i} \, \Delta_{j}}\biggr)^{5/2} \, 
\biggl(\frac{\rho_{i} \Delta_{j} + \rho_{j}\Delta_{i}}{2\rho_q}
\biggr)\,  \biggl(1 + \frac{5\rho^2_q}{2 - \rho^2_q} 
\, [ \Delta_{ij} \, - \, 1 ] \biggr) \, F_{ij}(p^2)
\en
where 
\eq
\Delta_{ij}  = \frac{2 \, \Delta_{i}^2 \, \Delta_{j}^2}
{\Delta_{i}^2 \, + \, \Delta_{j}^2} \,, \hspace*{1.5cm} 
\rho_{ij} = \frac{2  + 3 \rho^2_q}{(2  + 3 \rho^2_{i})^{1/2}
(2  + 3 \rho^2_{j})^{1/2}} 
\en
and $F_{ij}(p^2)$ is the meson-quark transition form factor 
normalized to unity at $p^2 = 0$: 
\eq
F_{ij}(p^2) =
\exp\bigg( - \frac{p^2 \, R^2}{4} \, \Delta_{ij} \biggr) \, 
\Biggl[1\,-\,\frac{p^2\,R^2}{2}\,\frac{\rho^2_q \, \Delta_{ij}^2} 
{2\,-\,6\rho^2_q\,+\,5\,\rho^2_q \,\Delta_{ij}} \Biggr] \,.
\en 

The gluon induced baryon mass shift $\Delta m_B^G$  
consists of the contribution of the gluon cloud (Fig.2a) and
the exchange (Fig.2b) diagram.  
As in the previous case we separate the gluon induced mass shifts 
$\Delta m_B^G$ into terms arising from the gluon cloud  
($\Delta m_B^{\rm E[C]}$ and $\Delta m_B^{\rm M[C]}$) and the gluon 
exchange ($\Delta m_B^{\rm E[E]}$ and $\Delta m_B^{\rm M[E]}$) 
corrections. A corresponding index is introduced in square brackets. 
In addition, the index ${\rm E}$ 
or ${\rm M}$ indicates contributions of Coulomb or transverse 
gluons, respectively:   
\eq\label{eqn:pcqmmassshifts_g}
&&\Delta m_B^G \, = \, \Delta m_B^{\rm G[C]} \, + \, 
\Delta m_B^{\rm G[E]}\,, \\
&&\nonumber\\
&&\Delta m_B^{\rm G[C]} \, = \,  \Delta m_B^{\rm E[C]} \, + \, 
\Delta m_B^{\rm M[C]}\,, \hspace*{1cm} 
\Delta m_B^{\rm G[E]} \, = \,  \Delta m_B^{\rm E[E]} \, + \, 
\Delta m_B^{\rm M[E]}\,, 
\en
where 
\eq  
\Delta m_B^{\rm I[C]} &=& 
d_{B}^{\rm I[C1]} \, \Sigma^{\rm I}_{q q} \, + \, 
d_{B}^{\rm I[C2]} \, \Sigma^{\rm I}_{s s} \,,\nonumber\\
&&\nonumber\\
\Delta m_B^{\rm I[E]} &=& 
d_{B}^{\rm I[E1]} \, \Sigma^{\rm I}_{q q} \, + \, 
d_{B}^{\rm I[E2]} \, \Sigma^{\rm I}_{q s} \, + \, 
d_{B}^{\rm I[E3]} \, \Sigma^{\rm I}_{s s} \,.\nonumber 
\en 
with ${\rm I = E, M}$. Here, $d_{B}^{\rm I[J]}$ are the recoupling 
coefficients which are listed in Table~\ref{tbl:dgcoeff} and 
$\Sigma^{{\rm E(M)}}_{q_1 q_2}$ are the gluon-loop integrals with 
\eq 
\Sigma^{\rm E}_{q_1 q_2} &=& \frac{1}{\pi} \int\limits_0^\infty dp \, 
\alpha_s(p^2) \, G_{\rm E}^{q_1}(p^2) \, G_{\rm E}^{q_2}(p^2) \, \\ 
\Sigma^{\rm M}_{q_1 q_2} &=& \frac{1}{\pi} \int\limits_0^\infty dp \, 
\alpha_s(p^2) \, G_{\rm M}^{q_1}(p^2) \, G_{\rm M}^{q_2}(p^2). \nonumber
\en
The charge and magnetic form factors $G_{\rm E}^i$ and $G_{\rm M}^i$ 
with $i=q,s$ of the constituent quarks are calculated at leading 
order (LO) or at tree level (without inclusion of the dressing by meson 
and gluon cloud corrections) using the solutions of the Dirac equation 
for nonstrange~(\ref{Gaussian_Ansatz}) and 
strange~(\ref{Gaussian_Ansatz_s}) quarks: 
\eq
G_{\rm E}^{i}(p^2) &=& \int\limits_0^\infty dx \, x^2 \, 
j_0(px) \biggl\{ \, [ g_0^i(x)]^2 + [f_0^i(x)]^2 \, \biggr\}  
\, = \,  \biggl(1 - \frac{\rho^2_i}{1+\frac{3}{2}\rho^2_i} 
\frac{p^2R^2_i}{4}\biggr) \, 
\exp\biggl(-\frac{p^2R^2_i}{4}\biggr) \,,\\ 
G_{\rm M}^{i}(p^2) &=& \int\limits_0^\infty dx \, x^2 \, 
j_1(px) \, \biggl\{  \, 2 \, g_0^i(x) \, f_0^i(x) \, \biggr\}  
\, = \, 2p \, \frac{\rho_i \,R_i}{1+\frac{3}{2}\rho^2_i} 
\, \exp\biggl(-\frac{p^2R^2_i}{4}\biggr) \,,\nonumber
\en 
where $g_0^i$ and $f_0^i$ are the upper and lower components of the 
Dirac spinors; $j_0$ and $j_1$ are the spherical Bessel functions. 
In Fig.3 we indicate the momentum dependence of the effective coupling 
as function of three-momentum squared (solid line). For comparison we 
also show the functions $\alpha_s(p^2)$ used in Ref.~\cite{Maris:1999nt} 
(long-dashed line) and the one referring to perturbative one-loop QCD 
(dotted line). The infrared part of $\alpha_s(p^2)$ in the present 
calculation is much smaller than the one used in 
Ref.~\cite{Maris:1999nt}. This again can be explained by the fact that 
in the infrared domain both gluon and meson degrees of freedom are 
strongly interacting with the quark fields. It leads in turn to 
a relative suppression of the effective running constant  
$\alpha_s(p^2)$ at low $p^2$ with respect to the original fit. 
The residual gluon contribution is therefore 
smaller than in the approach~\cite{Maris:1999nt} 
where mesonic degrees of freedom are not taken into account in 
the description of baryon structure. 

It is convenient to separate the value of the baryon mass $m_B$ into 
a term related to the chiral limit $\stackrel{0}{m}_B$ (defined in the 
limit $\hat m, m_s \to 0$ with  $\rho_s = \rho_q$ and $R_s = R_q$) and 
a term referring to the violation of chiral and unitary flavor symmetry 
$\delta m_B$: 
\eq
m_B \, = \, \stackrel{0}{m}_B \, + \, \delta m_B \,.  
\en 
Here  
\eq\label{Chiral_Res}
\stackrel{0}{m}_B &=& E_{3q}  
\, + \, \Delta \stackrel{\,\, 0 \, M}{m_B} 
\, + \, \Delta \stackrel{0 \, G}{m_B} \\
\Delta \stackrel{\,\, 0 \, M}{m_B} &=& 
\sum_{\Phi=\pi,K,\eta} \, \bigg\{ \,  
\Delta m_B^{\Phi {\rm [C]}}(0) \, + \, 
\Delta m_B^{\Phi {\rm [E]}}(0) 
\, \biggr\}\bigg|_{\rho_s = \rho_q\,, \, R_s = R_q} \,\,,\nonumber\\
\Delta \stackrel{0 \, G}{m_B} &=& \bigg\{ \, 
\Delta m_B^{\rm G[C]} \, + \, \Delta m_B^{\rm G[E]}
\,\biggr\}\bigg|_{\rho_s = \rho_q\,, \, R_s = R_q} \,\,.\nonumber
\en
Meson loops also contribute to 
$\stackrel{0}m_B$, since in the chiral limit quarks also interact with 
massless mesons. The separate contributions of the three-quark core
($E_{3q}$), meson ($\Delta \stackrel{\,\,0 \, M}{m_B}$) 
and gluon ($\Delta \stackrel{\,0 \, G}{m_B}$) cloud terms are: 
\eq
E_{3q} = 1612.4 \,\, {\rm MeV} 
\en 
for both flavor octet and decuplet baryons.
For the octet baryons we get   
\eq\label{MG_chir_8}
\Delta \stackrel{\!0 \, M}{m_{B^{8}}} &=& -254.6 \,\,{\rm MeV} 
-148.6 \,\,{\rm MeV} = -403.2 \,\,{\rm MeV} \\
\Delta \stackrel{\!\!0 \, G}{m_{B^{8}}} &=& 565.2 \,\,{\rm MeV} 
-946.0 \,\,{\rm MeV} = -380.8 \,\,{\rm MeV} \nonumber
\en 
while for the decuplet we have   
\eq\label{MG_chir_10}
\Delta \stackrel{\!\!\!0 \, M}{m_{B^{10}}} &=& -254.6 \,\,{\rm MeV} 
-42.5 \,\,{\rm MeV} = -297.1 \,\,{\rm MeV} \\
\Delta \stackrel{\!\!\!\!\!0 \, G}{m_{B^{10}}} &=& 565.2 \,\,{\rm MeV} 
-755.6 \,\,{\rm MeV} = -190.4 \,\,{\rm MeV}. \nonumber 
\en 
In Eqs.~(\ref{MG_chir_8}) and (\ref{MG_chir_10}) we indicate 
the contributions of the cloud (Figs.1a and 2a) and exchange 
(Figs.1b and 2b) diagrams derived in Eq.~(\ref{Chiral_Res}).  
Only the exchange diagrams of Figs.1b and 2b 
contribute to the octet-decuplet mass splitting in the 
chiral limit: 
\eq
\delta = \stackrel{0}{m}_{B^{10}} - 
\stackrel{0}{m}_{B^{8}} = \delta^M + \delta^G 
\en 
where $\delta^M = 106.1$ MeV and $\delta^G = 190.4$ MeV are the
partial meson 
and gluon contributions. 
  
The octet baryon mass ($\stackrel{0}{m}_{B^{8}}$) and a finite 
decuplet-octet baryon mass difference, both defined in the chiral limit,
are the natural 
ingredients (or input parameters) of chiral effective Lagrangians 
(see, e.g. 
Refs.~\cite{Gasser:1987rb,Jenkins:1992pi,Borasoy:1996bx,Durand:1997ya,Becher:1999he,Ellis:1999jt,Young:2002cj,Procura:2003ig}). 
Our numerical values for these quantities are: 
\eq\label{Chiral_Res2}
\stackrel{0}{m}_{B^{8}} &=& 828.5 \,\, {\rm MeV}\,, 
\hspace*{1.3cm} B^{8} = N, \Lambda, \Sigma, \Xi   \\
\stackrel{0}{m}_{B^{10}} &=& 1124.9 \,\, {\rm MeV}\,, 
\hspace*{1cm} B^{10} = \Delta, \Sigma^\ast, \Xi^\ast, \Omega 
\nonumber\\
\delta &=& 296.4 \,\, {\rm MeV} \approx 300 \,\, {\rm MeV} 
\,.\nonumber
\en 
Our value for $\delta$ is close to the experimental 
result for the $\Delta-N$ mass splitting. Inclusion of symmetry breaking 
corrections does not affect the value for 
$\stackrel{0}{m}_\Delta - \stackrel{0}{m}_N$ evaluated in the chiral 
limit. For comparison we quote the parameters of the average octet 
baryon mass $\stackrel{0}{m}_{B^{8}}$ and the mass difference $\delta$ 
as predicted or adjusted in other effective chiral approaches. The value 
of $\stackrel{0}{m}_{B^{8}}$ was estimated in heavy baryon chiral 
perturbation theory~\cite{Borasoy:1996bx} with 
$\stackrel{0}{m}_{B^{8}}= 770 \pm 110$ MeV. In chiral perturbation 
theory with infrared regularization~\cite{Ellis:1999jt}
$\stackrel{0}{m}_{B^{8}}$ decreases from 733 MeV at third order of the 
chiral expansion to 653 MeV at fourth order. 
In Ref.~\cite{Durand:1997ya} the input parameter $\delta = 300$ MeV, 
which is very close to ours, was used in the description of baryon 
magnetic moments. A detailed analysis of baryon (nucleon) masses using 
a chiral extrapolation of lattice data was done in 
Refs.~\cite{Young:2002cj,Procura:2003ig}. In Ref.~\cite{Young:2002cj} 
the meson-loop contribution to the $\Delta-N$ mass splitting for 
different values of $M_\pi^2$ (including the chiral limit $M_\pi^2=0$ 
and the physical point) was studied in both schemes (quenched and full) 
of lattice QCD. The pion cloud contributes not more than one third to 
the observed $\Delta-N$ mass splitting. For a specific choice of the 
regulator parameter meson loops result in about 50 MeV in the
full scheme of lattice QCD. The remaining contribution to the splitting
is probably attributed to short-range effects, 
like gluon exchange. Below we show that the lattice finding corresponds
to the situation derived in our model: the partial contributions of 
meson and gluon cloud to $\delta$ are $1/3$ and $2/3$, respectively. 
In Ref.~\cite{Procura:2003ig} a value of $\stackrel{0}{m}_{N} \simeq 880$~MeV 
was deduced with baryon chiral perturbation theory up to order $p^4$. 

In Table~\ref{tbl:ma_shifts} we give the results for the
partial and total mass 
shifts of the octet and decuplet baryons for our set 
of parameters: $\rho_q = \sqrt{2/13}$, $\rho_s = 0.354$, 
$R_q = 0.6$ fm and $R_s = 0.58 $ fm.   
Table~\ref{tbl:mb_shifts} contains the results for the case when 
the parameters of nonstrange and strange quarks are degenerate with  
$\rho_q = \rho_s = \sqrt{2/13}$ and $R_q = R_s = 0.6$ fm. 
The meson cloud provides a significant downward shift,
e.g., for nucleon it is about $- 300$ MeV. 
Thereby the dominant contribution is due to pion loops with
$\sim - 260$ MeV. The two sets of results, with and without inclusion
of flavor symmetry breaking, presented in
Tables~\ref{tbl:ma_shifts} and~\ref{tbl:mb_shifts} display no big
difference.
The results differ slightly in particular
for hyperons due to the modification of the kaon cloud. 
Meson loop effects also provide considerable 
mass splittings. 
For example, they contribute 110 MeV to the splitting $m_{\Xi} - m_{N}$, 
which is of considerable size. However, the meson induced mass shifts 
are not sufficient to explain the full observed mass 
splittings. For example, meson effect yield only 100 MeV to the mass 
difference $m_{\Delta} - m_{N}$, and also only 40 MeV to
$m_{\Sigma} - m_{\Lambda}$. 
The observed splittings are about 300 MeV and 75 MeV, respectively.  
An additional mechanism, like effective gluon exchange,
is needed to explain the 
observed splittings. The important ingredients to partially reproduce
the mass shifts are at this level already taken into account:
breaking of unitary flavor symmetry, 
meson cloud effects and finite current quark masses (especially for 
the strange quark). 

In a next step we also include the quantum corrections of the gluon cloud
to the baryon masses. The numerical 
results for the mass shifts induced by the gluon corrections are given
in Tables~\ref{tbl:ga_shifts} 
and~\ref{tbl:gb_shifts}. Again, we present our predictions for the two 
sets of parameters $\rho_q = \sqrt{2/13}$, $\rho_s = 0.354$, 
$R_q = 0.6$ fm and $R_s = 0.58 $ fm (Table~\ref{tbl:ga_shifts}) and 
$\rho_q = \rho_s = \sqrt{2/13}$ and $R_q = R_s = 0.6$ fm 
(Table~\ref{tbl:gb_shifts}). In the "symmetric" case, 
where the bare parameters of the quark wave functions are identical, 
gluon cloud corrections to the baryon octet and decuplet mass shifts 
are degenerate. 

Combining all effects we can reasonably reproduce all the observed mass 
differences. In Table~\ref{tbl:mass_a} we give the full results for the 
ground-state baryon masses indicating all mechanisms (bare mass, 
mass shifts induced by the current quark mass, meson and gluon cloud) 
for the best choice of free parameters. For the experimental 
data we quote the masses of $m_P, m_{\Lambda^0}, m_{\Sigma^+}, 
m_{\Xi^-}, m_{\Delta^+}, m_{\Sigma^{\ast\,+}}, m_{\Xi^{\ast\,-}}$ 
and $m_{\Omega^-}$.  For completeness, in Table~\ref{tbl:mass_b}
we also list our results for the symmetric set of free parameters. 
Obviously, in the latter case the description of the hyperon mass 
spectrum deteriorates with respect to the "asymmetric case" where the 
mechanism of $SU(3)$ flavor symmetry breaking is included. In the 
following we only discuss the full results presented in 
Table~\ref{tbl:mass_a}. The predicted mass of the $\Sigma$ 
baryon is heavier than that of the $\Lambda$ hyperon consistent with
the experimental observation, although the obtained difference is 
a little smaller than the data. The difference is generated both by 
meson and gluon induced mass shifts, but the mesonic effect is about 
40 MeV and hence more important than the one due to the gluon. 
A slightly more "asymmetric parameter" set results in a larger 
gluon induced mass difference and hence a better fit to the
$\Sigma - \Lambda$ mass splitting, but the overall mass prediction gets 
worse. Next we illustrate the validity of our model in the context of 
the Gell-Mann-Okubo mass relations (\ref{GMO}). Both the 
meson and gluon induced mass shifts well satisfy the relations 
separately, based on our construction of $SU(6)$ baryon wave 
functions; hence the predictions for the total masses also satisfy
these relations.
Moreover, our model generates small violations of the relations
which are qualitatively consistent with data, i.e. 
L.H.S. $>$ R.H.S. in the octet and L.H.S. $>$ Center $>$ R.H.S. 
in the decuplet.
These phenomenologically consistent violations are provided by the meson 
cloud, while the gluon corrections give opposite signs. Finally, 
we comment on the $SU(3)$ breaking effects on the meson-nucleon 
sigma terms, which are fully discussed in Ref. ~\cite{Inoue:2003bk}.
We list the results in Table~\ref{tbl:sigma}, where the additional
effect is found to be negligibly small. For example, the obtained 
$\pi N$ sigma term of 54.6 MeV is merely 0.1 MeV smaller than in the case
of the symmetric parameter set. This is naturally to be expected since 
the nucleon does not contain strange valence
quarks and hence only kaon loop diagrams are affected by 
the flavor symmetry breaking. Again, the "asymmetric parameter" set is 
more important when considering hyperons.
Therefore, the results for the meson-nucleon sigma terms indicated in a
recent paper~\cite{Inoue:2003bk}, where the "symmetric parameter" set is 
employed, are essentially reasonable.

\section{Summary}
\label{Sect_summary} 
We have studied the mass spectrum of the ground-state baryons
in the perturbative 
chiral quark model. We have extended the model by including
quantum corrections of the gluon and by setting up a more realistic
mechanism for $SU(3)$ flavor symmetry violation.  
Both, meson and gluon induced mass shifts are important ingredients
to quantitatively reproduce the absolute baryon mass spectrum and
the resulting mass splittings. The mass splitting between the octet 
and the decuplet baryons is dominantly generated by the gluon 
mechanism. In particular, the $\Delta-N$ mass difference
is saturated by $1/3$ by the meson cloud, while gluon corrections
contribute about $2/3$. This finding 
is consistent with a recent lattice simulation~\cite{Young:2002cj},  
where a sizable nucleon-$\Delta$ mass  
splitting is even observed in the quenched approximation.
For the mass scales within each multiplet the meson mechanism is found
to be the most important effect, as demonstrated for example
for the $\Sigma - \Lambda$ mass difference.
In addition, the meson mechanism is qualitatively consistent with the 
small violations of the Gell-Mann-Okubo mass relations.

\vspace*{1.5cm}

{\bf Acknowledgments}

\noindent
This work was supported by the Deutsche Forschungsgemeinschaft (DFG) 
under contracts FA67/25-3 and GRK683.


\begin{table}
\caption{Meson recoupling coefficients for the octet and 
decuplet baryons in units of $1/400$.} 
\label{tbl:dcoeff}
\begin{center}
\def\arraystretch{1.15}
\begin{tabular}{ccccccccccc}
$d^{\Phi{\rm [J]}}_{B}$ & ${^{\displaystyle{\pi[\rm C]}}}$ 
                        & ${^{\displaystyle{K[\rm C1]}}}$  
                        & ${^{\displaystyle{K[\rm C2]}}}$ 
                        & ${^{\displaystyle{\eta[\rm C1]}}}$ 
                        & ${^{\displaystyle{\eta[\rm C2]}}}$  
                        & ${^{\displaystyle{\pi[\rm E]}}}$ 
                        & ${^{\displaystyle{K[\rm E]}}}$  
                        & ${^{\displaystyle{\eta[\rm E1]}}}$
                        & ${^{\displaystyle{\eta[\rm E2]}}}$ 
                        & ${^{\displaystyle{\eta[\rm E3]}}}$\\
 \hline 
 $N$            & 81 & 54 & 0  & 9 & 0  & 90 & 0  & - 6 & 0  & 0 \\
 \hline 
 ${\Lambda}$    & 54 & 36 & 36 & 6 & 12 & 54 & 36 & - 6 & 0  & 0 \\
 \hline 
 ${\Sigma}$     & 54 & 36 & 36 & 6 & 12 & 6  & 60 &   2 & 16 & 0 \\
 \hline
 ${\Xi}$        & 27 & 18 & 72 & 3 & 24 & 0  & 60 &   0 & 16 & 8 \\
 \hline
 ${\Delta}$     & 81 & 54 & 0  & 9 & 0  & 18 & 0  &   6 & 0  & 0 \\
 \hline 
 ${\Sigma^\ast}$& 54 & 36 & 36 & 6 & 12 & 6  & 24 &   2 & -8 & 0 \\ 
 \hline
 ${\Xi^\ast}$   & 27 & 18 & 72 & 3 & 24 & 0  & 24 &   0 & -8 & 8 \\ 
 \hline
 ${\Omega}$     & 0  & 0  &108 & 0 & 36 & 0  & 0  &   0 & 0  & 24\\ 
\end{tabular}
\normalsize
\end{center}
\end{table}

\begin{table}
\caption{Gluon recoupling coefficients for the octet and 
decuplet baryons in units of $1/9$.} 
\label{tbl:dgcoeff}
\begin{center}
\def\arraystretch{1.15}
\begin{tabular}{ccccccccccc}
$d^{\rm I [J]}_{B}$ & ${^{\displaystyle{\rm E[C1]}}}$ 
                    & ${^{\displaystyle{\rm E[C2]}}}$ 
                    & ${^{\displaystyle{\rm M[C1]}}}$  
                    & ${^{\displaystyle{\rm M[C2]}}}$  
	            & ${^{\displaystyle{\rm E[E1]}}}$ 
                    & ${^{\displaystyle{\rm E[E2]}}}$ 
		    & ${^{\displaystyle{\rm E[E3]}}}$
                    & ${^{\displaystyle{\rm M[E1]}}}$  
                    & ${^{\displaystyle{\rm M[E2]}}}$  
                    & ${^{\displaystyle{\rm M[E3]}}}$ \\ 
 \hline 
 $N$            & 36  & 0   & -72 &   0 & -36 &   0 &   0 & -24  
                &  0  &  0 \\
 \hline 
 ${\Lambda}$    & 24  & 12  & -48 & -24 & -12 & -24 &   0 & -24  
                &   0  &  0 \\
 \hline 
 ${\Sigma}$     & 24  & 12  & -48 & -24 & -12 & -24 &   0 &   8  
                & -32  &  0 \\
 \hline
 ${\Xi}$        & 12  & 24  & -24 & -48 &   0 & -24 & -12 &   0  
                & -32  &  8 \\
 \hline
 ${\Delta}$     & 36  & 0   & -72 &   0 & -36 &   0 &   0 &  24  
                &   0  &  0 \\
 \hline 
 ${\Sigma^\ast}$& 24  & 12  & -48 & -24 & -12 & -24 &   0 &   8  
                &  16  &  0 \\
 \hline
 ${\Xi^\ast}$   & 12  & 24  & -24 & -48 &   0 & -24 & -12 &   0  
                &  16  &  8 \\
 \hline
 ${\Omega}$     & 0   & 36  &  0  & -72 &   0 &   0 & -36 &   0  
                &   0  & 24 
\end{tabular}
\normalsize
\end{center}
\end{table}

\begin{table}
\caption{Baryon mass shifts (in units of MeV) induced 
by the meson cloud for the parameter set $\rho_q = \sqrt{2/13}$, 
$\rho_s = 0.354$, $R_q = 0.6$ fm and $R_s = 0.58 $ fm.} 
\label{tbl:ma_shifts}
\begin{center}
\def\arraystretch{1.1}
\begin{tabular}{crrrrrrr}
$\Delta m_B^{\Phi[\rm J]}$  
& ${^{\displaystyle{\pi[\rm C]}}}$ 
& ${^{\displaystyle{K[\rm C]}}}$ 
& ${^{\displaystyle{\eta[\rm C]}}}$  
& ${^{\displaystyle{\pi[\rm E]}}}$ 
& ${^{\displaystyle{K[\rm E]}}}$ 
& ${^{\displaystyle{\eta[\rm E]}}}$  
& Total\\
\hline 
$N$           & $-125.2$ & $-40.0$ &  $-6.2$ & $-139.1$ &     $0$ 
              &  $ 4.1$ & $-306.5$
 \\
 \hline 
$\Lambda$     &  $-83.5$ & $-53.4$ & $-11.5$ &  $-83.5$ & $-26.7$ 
              &  $ 4.1$ & $-254.5$
 \\
 \hline
$\Sigma$      &  $-83.5$ & $-53.4$ & $-11.5$ &   $-9.3$ & $-44.5$ 
              & $-11.8$ & $-214.0$
 \\
 \hline
$\Xi$         &  $-41.7$ & $-66.7$ & $-16.9$ &     $0$  & $-44.5$ 
              & $-15.3$ & $-185.1$
 \\
 \hline
$\Delta$      & $-125.2$ & $-40.0$ &  $-6.2$ &  $-27.8$ &     $0$ 
              &  $-4.1$ & $-203.3$
 \\
 \hline 
$\Sigma^\ast$ &  $-83.5$ & $-53.4$ & $-11.5$ &   $-9.3$ & $-17.8$ 
              &   $3.8$ & $-171.7$
 \\
 \hline
$\Xi^\ast$    &  $-41.7$ & $-66.7$ & $-16.9$ &      $0$ & $-17.8$ 
              &   $0.3$ & $-142.8$
 \\
 \hline
$\Omega$      &     $0$  & $-80.1$ & $-22.2$ &      $0$ &     $0$ 
              & $-14.8$ & $-117.1$
\end{tabular}
\end{center}
\end{table}

\begin{table}
\caption{Baryon mass shifts (in units of MeV) induced by the meson 
cloud for the parameter set $\rho_q = \rho_s = \sqrt{2/13}$ and 
$R_q = R_s = 0.6$ fm.} 
\label{tbl:mb_shifts}
\begin{center}
\def\arraystretch{1.1}
\begin{tabular}{crrrrrrr}
$\Delta m_B^{\Phi[\rm J]}$  
& ${^{\displaystyle{\pi[\rm C]}}}$ 
& ${^{\displaystyle{K[\rm C]}}}$ 
& ${^{\displaystyle{\eta[\rm C]}}}$  
& ${^{\displaystyle{\pi[\rm E]}}}$ 
& ${^{\displaystyle{K[\rm E]}}}$ 
& ${^{\displaystyle{\eta[\rm E]}}}$  
& Total\\
\hline 
$N$           & $-125.2$ & $-42.3$ &  $-6.2$ & $-139.1$ &     $0$ 
              &  $ 4.1$  & $-308.7$
 \\
 \hline 
$\Lambda$     &  $-83.5$ & $-56.4$ & $-12.3$ &  $-83.5$ & $-28.2$ 
              &  $ 4.1$  & $-259.8$
 \\
 \hline
$\Sigma$      &  $-83.5$ & $-56.4$ & $-12.3$ &   $-9.3$ & $-47.0$ 
              & $-12.3$  & $-220.8$
 \\
 \hline
$\Xi$         &  $-41.7$ & $-70.6$ & $-18.5$ &      $0$ & $-47.0$ 
              & $-16.4$  & $-194.2$
 \\
 \hline
$\Delta$      & $-125.2$ & $-42.3$ &  $-6.2$ &  $-27.8$ &     $0$ 
              &  $-4.1$  & $-205.6$
 \\
 \hline 
$\Sigma^\ast$ &  $-83.5$ & $-56.4$ & $-12.3$ &   $-9.3$ & $-18.8$ 
              &   $4.1$  & $-176.2$
 \\
 \hline
$\Xi^\ast$    &  $-41.7$ & $-70.6$ & $-18.5$ &      $0$ & $-18.8$ 
              &     $0$  & $-149.6$
 \\
 \hline
$\Omega$      &      $0$ & $-84.7$ & $-24.7$ &      $0$ &     $0$ 
              & $-16.4$  & $-125.8$
\end{tabular}
\end{center}
\end{table}

\begin{table}
\caption{Baryon mass shifts (in units of MeV)
induced by the gluon cloud for the parameter set
$\rho_q = \sqrt{2/13}$, $\rho_s = 0.354$, 
$R_q = 0.6$ fm and $R_s = 0.58 $ fm.} 
\label{tbl:ga_shifts}
\begin{center}
\def\arraystretch{1.1}
\begin{tabular}{crrrrr}
$\Delta m_B^{\rm I[J]}$  
& ${^{\displaystyle{\rm E[C]}}}$ 
& ${^{\displaystyle{\rm M[C]}}}$ 
& ${^{\displaystyle{\rm E[E]}}}$ 
& ${^{\displaystyle{\rm M[E]}}}$ 
&  Total\\
 \hline 
 $N$           & $850.8$ & $-285.6$ & $-850.8$ & $-95.2$ & $-380.8$
 \\
 \hline 
 $\Lambda$     & $870.9$ & $-274.2$ & $-870.4$ & $-95.2$ & $-368.9$
 \\
 \hline
 $\Sigma$      & $870.9$ & $-274.2$ & $-870.4$ & $-87.3$ & $-361.0$
 \\
 \hline
 $\Xi$         & $890.9$ & $-262.7$ & $-890.4$ & $-91.1$ & $-353.3$
 \\
 \hline
 $\Delta$      & $850.8$ & $-285.7$ & $-850.8$ &  $95.2$ & $-190.4$
 \\
 \hline 
 $\Sigma^\ast$ & $870.9$ & $-274.2$ & $-870.4$ &  $91.2$ & $-182.4$
 \\
 \hline
 $\Xi^\ast$    & $890.9$ & $-262.7$ & $-890.4$ &  $87.4$ & $-174.8$
 \\
 \hline
 $\Omega$      & $910.9$ & $-251.2$ & $-910.9$ &  $83.7$ & $-167.5$  
\end{tabular}
\end{center}
\end{table}

\begin{table}
\caption{Baryon mass shifts (in units of MeV)
induced by the gluon cloud for the parameter set
$\rho_q = \rho_s = \sqrt{2/13}$ and $R_q = R_s = 0.6$ fm.}  
\label{tbl:gb_shifts}
\begin{center}
\def\arraystretch{1.1}
\begin{tabular}{crrrrr}
$\Delta m_B^{\rm I[J]}$  
& ${^{\displaystyle{\rm E[C]}}}$ 
& ${^{\displaystyle{\rm M[C]}}}$ 
& ${^{\displaystyle{\rm E[E]}}}$ 
& ${^{\displaystyle{\rm M[E]}}}$ 
&  Total \\
 \hline 
$N, \Lambda, \Sigma, \Xi$  
& $850.8$ & $-285.6$ & $-850.8$ & $-95.2$ & $-380.8$
\\
\hline
$\Delta, \Sigma^\ast, \Xi^\ast, \Omega$  
& $850.8$ & $-285.6$ & $-850.8$ & $95.2$ & $-190.4$
\end{tabular}
\end{center}
\end{table}

\newpage

\begin{table}
\caption{Full model results for the baryon mass spectrum 
(in units of MeV) for the set of parameters $\rho_q = \sqrt{2/13}$, 
$\rho_s = 0.354$, $R_q = 0.6$ fm and $R_s = 0.58 $ fm.} 
\label{tbl:mass_a}
\begin{center}
\def\arraystretch{1.1} 
\begin{tabular}{lrrrrrr}
       &  Core     & Quark   &  Mesons & Gluons   &  Total & Data
 \\
 \hline
 $m_N$             &  $1612.4$ &  $13.1$  & $-306.4$ &  $-380.8$ 
       &   $938.3$ &  938.3
 \\
 \hline 
 $m_{\Lambda}$     &  $1612.4$ & $128.3$  & $-254.4$ &  $-368.9$ 
       &  $1117.4$ & 1115.7
 \\
 \hline
 $m_{\Sigma}$      &  $1612.4$ & $128.3$  & $-213.9$ &  $-361.0$ 
       &  $1165.9$ & 1189.4
 \\
 \hline
 $m_{\Xi}$         &  $1612.4$ & $243.5$  & $-185.1$ &  $-353.3$ 
       &  $1317.5$ & 1321.3
 \\ 
 \hline
 $m_{\Delta}$      &  $1612.4$ &  $13.1$  & $-203.4$ &  $-190.4$ 
       &  $1231.8$ & 1231.6
 \\
 \hline 
 $m_{\Sigma^\ast}$ &  $1612.4$ & $128.3$  & $-171.6$ &  $-182.4$ 
       &  $1386.7$ & 1382.8
 \\
 \hline
 $m_{\Xi^\ast}$    &  $1612.4$ & $243.5$  & $-142.9$ &  $-174.8$ 
       &  $1538.3$ & 1535.0
 \\
 \hline
 $m_{\Omega}$      &  $1612.4$ & $358.6$  & $-117.1$ &  $-167.5$ 
       &  $1686.5$ & 1672.5
\end{tabular}
\end{center}
\end{table}

\begin{table}
\caption{Full model results for the baryon mass spectrum 
(in units of MeV) for the set of parameters 
$\rho_q = \rho_s = \sqrt{2/13}$ and $R_q = R_s = 0.6$ fm.}  
\label{tbl:mass_b}
\begin{center}
\def\arraystretch{1.1} 
\begin{tabular}{lrrrrrr}
      &  Core   & Quark    &  Mesons    & Gluons &  Total   & Data
 \\
 \hline
 $m_N$             &  $1612.4$ &  $13.1$  &  $-308.8$  &  $-380.8$ 
      &   $935.9$  &  938.3
 \\
 \hline 
 $m_{\Lambda}$     &  $1612.4$ & $118.1$  &  $-259.8$  &  $-380.8$ 
      &  $1089.9$  & 1115.7
 \\
 \hline
 $m_{\Sigma}$      &  $1612.4$ & $118.1$  &  $-220.9$  &  $-380.8$ 
      &  $1128.8$  & 1189.4
 \\
 \hline
 $m_{\Xi}$         &  $1612.4$ & $223.1$  &  $-194.3$  &  $-380.8$ 
      &  $1260.4$  & 1321.3
 \\ 
 \hline
 $m_{\Delta}$      &  $1612.4$ &  $13.1$  &  $-205.7$  &  $-190.4$ 
      &  $1229.4$  & 1231.6
 \\
 \hline 
 $m_{\Sigma^\ast}$ &  $1612.4$ & $118.1$  &  $-176.2$  &  $-190.4$ 
      &  $1363.9$  & 1382.8
 \\
 \hline
 $m_{\Xi^\ast}$    &  $1612.4$ & $223.1$  &  $-149.6$  &  $-190.4$ 
      &  $1495.5$  & 1535.0
 \\
 \hline
 $m_{\Omega}$      &  $1612.4$ & $328.2$  &  $-125.8$  &  $-190.4$ 
      &  $1624.4$  & 1672.5
\end{tabular}
\end{center}
\end{table}

\begin{table}
\caption{Meson-nucleon sigma terms (in units of MeV) for the set of
parameter (I) $\rho_q = \sqrt{2/13}$, $\rho_s = 0.354$, $R_q = 0.6$ fm 
and $R_s = 0.58 $ fm,
and (II) $\rho_q = \rho_s = \sqrt{2/13}$ and $R_q = R_s = 0.6$ fm }
\label{tbl:sigma}
\begin{center}
\def\arraystretch{1.1}
\begin{tabular}{ccccc}
 Set & $\sigma_{\pi N}$ & $\sigma_{K N}^u$ & $\sigma_{K N}^d$ 
 & $\sigma_{\eta N}$ \\ \hline
 I  & 54.6 & 417.1 & 351.7 & 93.0  \\         
 II & 54.7 & 419.2 & 353.4 & 96.3                     
\end{tabular}
\end{center}
\end{table}

\newpage
\begin{figure}[t]
\noindent Fig.1: Meson loop diagrams contributing to the baryon energy 
shift: 

\noindent meson cloud (1a) and meson exchange diagram (1b). 

\vspace*{1.5cm}

\noindent Fig.2: Gluon loop diagrams contributing to the baryon energy 
shift: 

\noindent gluon cloud (2a) and gluon exchange diagram (2b). 

\vspace*{1.5cm}

\noindent Fig.3: Effective running couplings $\alpha_s(p^2)$: 
i) used in the present study (solid line),  
ii) used in ref.~\cite{Maris:1999nt} with the original parameterization 
(long-dashed line) and iii) used in perturbative QCD (dotted line). 
\end{figure}

\newpage 

\vspace*{-5cm}

\begin{figure}
\centering{\
\epsfig{figure=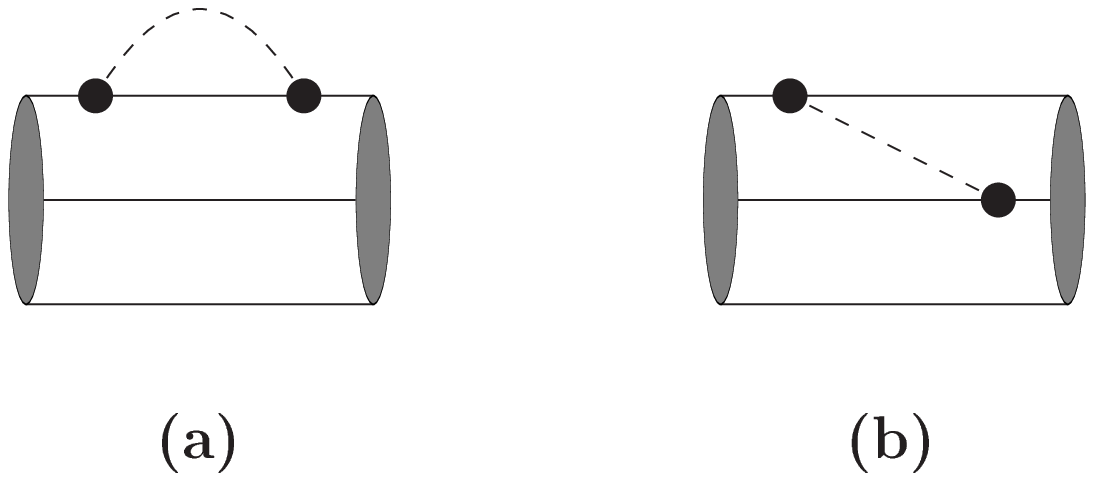,height=21cm}}

\vspace*{-8cm}

\centerline{\bf Fig.1}

\vspace*{-3cm}

\centering{\
\epsfig{figure=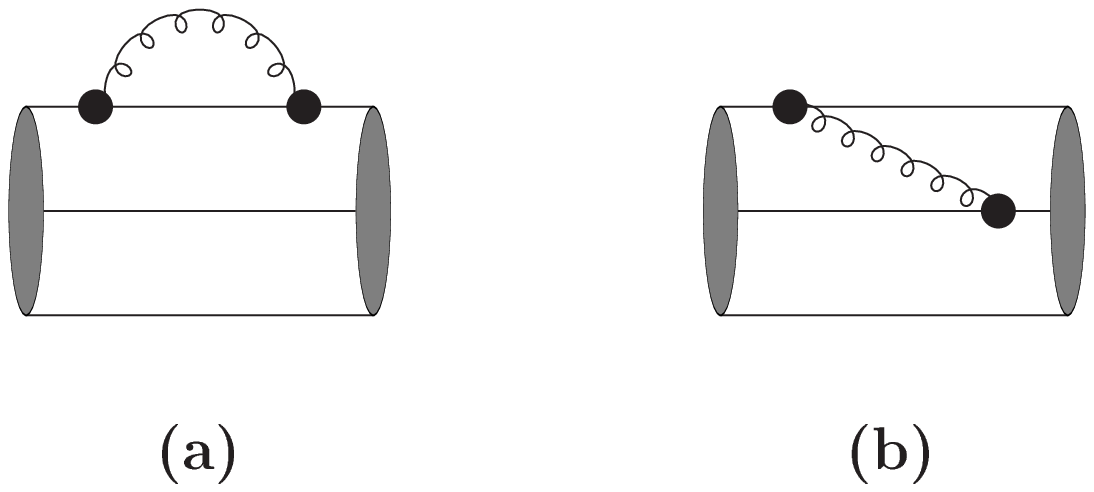,height=21cm}}

\vspace*{-8cm}

\centerline{\bf Fig.2}

\end{figure}

\newpage

\begin{figure}

\centering{\
\epsfig{figure=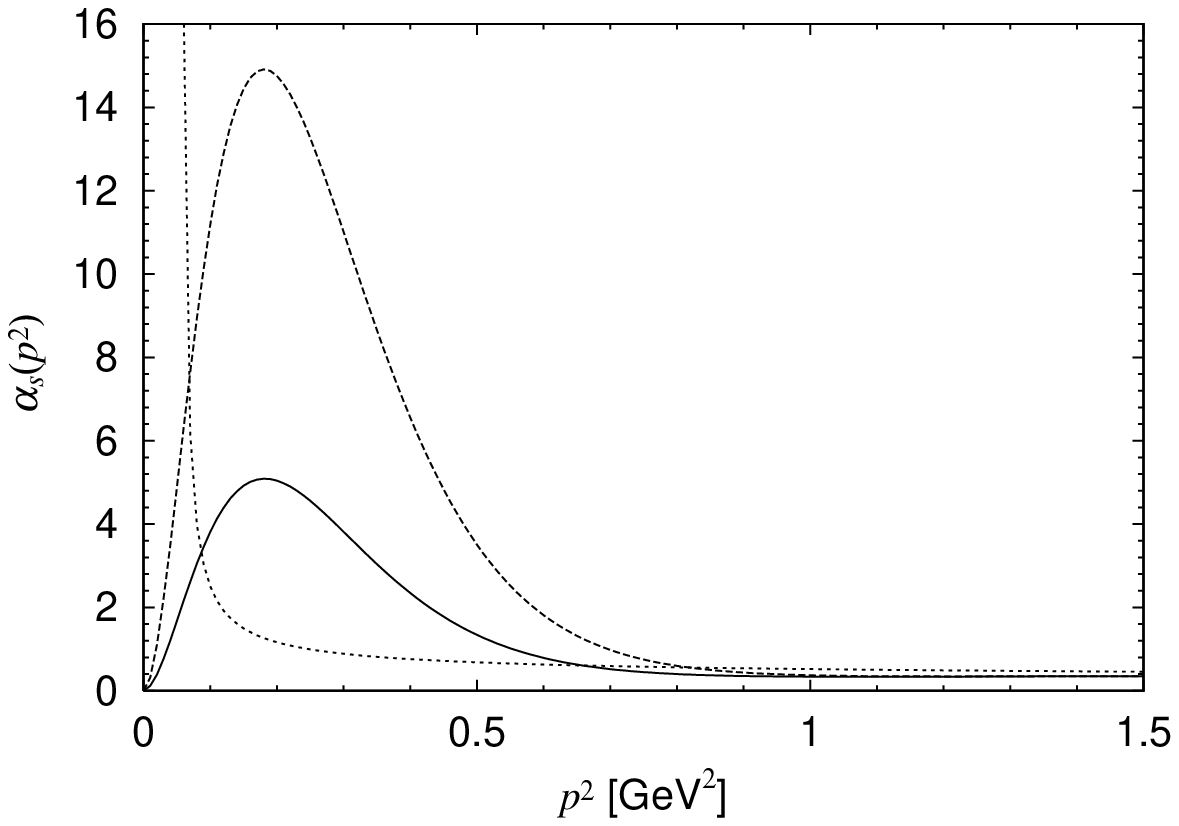,height=10cm}}

\centerline{\bf Fig.3}

\end{figure}

\end{document}